# Effect of wave versus particle phonon nature in thermal transport through nanostructures

Dhritiman Chakraborty[1*], Hossein Karamitaheri[2], Laura de Sousa Oliveira[1] and Neophytos Neophytou[1]

[1]School of Engineering, University of Warwick, Coventry, CV4 7AL, UK
[2]Department of Electrical Engineering, University of Kashan, Kashan, Iran
[*] D.Chakraborty@warwick.ac.uk

## Abstract

Comprehensive understanding of thermal transport in nanostructured materials needs large scale simulations bridging length scales dictated by different physics related to the wave versus particle nature of phonons. Yet, available computational approaches implicitly treat phonons as either just waves or as particles. In this work, using a full wave-based Non-Equilibrium Green's Function (NEGF) method, and a particle-based ray-tracing Monte Carlo (MC) approach, we investigate the qualitative differences in the wave and particle-based phonon transport at the vicinity of nanoscale features. For the simple example of a nanoporous geometry, we show that phonon transmission agrees very well for both methods with an error margin of ± 15%, across phonon wavelengths even for features with sizes down to 3-4 nm. For cases where phonons need to squeeze in smaller regions to propagate, we find that MC underestimates the transmission of long wavelength phonons whereas wave treatment within NEGF indicates that those long wavelength phonons can propagate more easily. We also find that particle-based simulation methods are somewhat more sensitive to structural variations compared to the wave-based NEGF method. The insight extracted from comparing wave and particle methods can be used to provide a better and more complete understanding of phonon transport in nanomaterials.

# Graphical Abstract

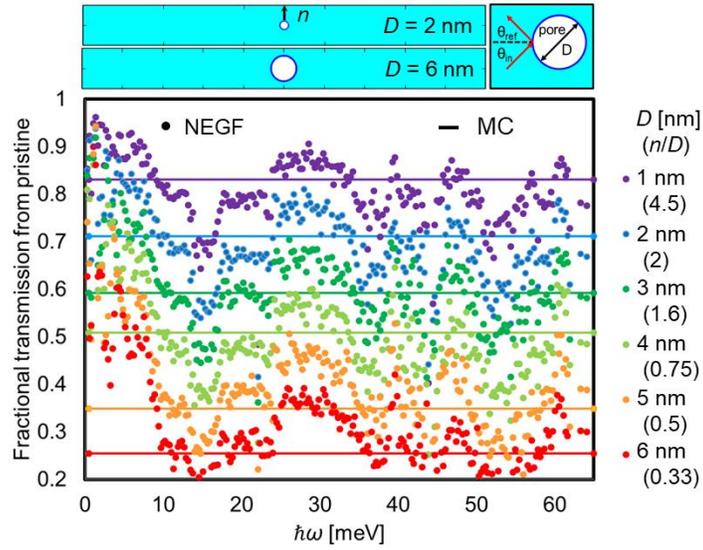

Figure caption: Geometry schematics for typical structures studied. NEGF fractional transmission (dots) and MC fractional transmission (lines) for structures with pore diameters $D = 1$ nm (purple), $D = 2$ nm (blue), $D = 3$ nm (dark green), $D = 4$ nm (light green), $D = 5$ nm (orange), $D = 6$ nm (red) respectively. The pore diameters ($D$) and the corresponding neck to diameter ($n/D$) ratios are indicated on the right axis.

# I. Introduction

Nanostructuring has magnified our ability to understand and control phonon transport in nanomaterials, giving rise to novel applications. Phonon waveguides, which can be used for nanoelectromechanical systems (NEMS) [1], phonovoltaics, which uses high energy phonons to generate power, similar to photovoltaics [2, 3], heat management, and primarily thermoelectrics, which is a way to generate power from temperature gradients [4], are a few emerging technologies that depend on heat flow at the nanoscale. Nanostructuring of materials has demonstratively yielded extremely low thermal conductivity ($\kappa$), even below the amorphous limit of materials, resulting in enhanced thermoelectric efficiency [5, 6]. Particularly for porous silicon materials, novel single-crystalline membranes with nanoscopic pores give reproducibly low $\kappa$ around 1-2 $\text{Wm}^{-1}\,\text{K}^{-1}$ [7 - 9].

Optimizing and controlling heat flow prerequisites a theoretical understanding of phonon transport in these materials. This requires methods that bridge through various length scales which are dictated by different physics related to the wave versus particle nature of phonons. Yet, conventional computational methods consider these approaches (i.e., waves *vs* particles) independently, and whether and under what circumstances wave effects and coherence have significant influence in phonon transport is still an open question [10 - 13]. Classical methods based on the Boltzmann Transport Equation (BTE) treat phonons as incoherent particles [12] and have been used very successfully to describe thermal transport properties in silicon across various nanostructures such as nanowires [14 -16], thin films [17 - 19], nanoporous materials [20 - 30], polycrystalline materials [31 - 36], nanocomposites [37, 38], silicon-on-insulator devices [39], and corrugated structures [40 - 43]. These methods scale very well computationally and allow flexibility in the treatment of scattering and geometrical features (Monte Carlo for example), and are commonly employed to simulate thermal transport in nanostructures.

However, these classical methods do not consider phonon wave effects that could be important at the nanoscale. For example, adding periodic pores in a thin film system to create a nanomesh [6, 8, 9, 20, 44] introduces a secondary artificial periodicity to the original lattice, potentially modifying the phonon dispersion relations to form a "phononic crystal", analogous to the better-known photonic crystals [45]. Changes in phonon group velocities, band gaps, the density of states, and thermal conductivity that arise due to wave interactions are referred to as *coherent effects*. It is also suggested that such effects can lead to Anderson localization, mode conversion, or Rayleigh waves and drastically reduce $\kappa$ [46, 47]. However, experimental results are still inconclusive on the relative importance of (wave-based) coherence effects versus (particle-based) boundary scattering effects [12, 13] at length scales below 100 nm [6 - 11, 48 - 51]. Theoretical investigations can help improve the understanding of phonon transport at these length scales, but investigations which include both wave-based and also particle-based methods are scarce. Fully understanding the qualitative and quantitative effects of such geometries on thermal transport and phonon transmission functions would allow the design of more efficient materials and devices for thermoelectric, heat management, and other phononic applications.

In this work, we attempt to shed some light on the difference in phonon transport through structures with nanoscale feature sizes when phonons are purely waves, versus being purely particles. We compare phonon transport in a simple porous silicon geometry using a full wave approach based on the atomistic Non-Equilibrium Green's Function (NEGF) method, and a particle–based approach using the ray-tracing Monte Carlo (MC) method, and quantify how much of the wave effects MC captures. Our motivation resides in the fact that MC is a truly valuable tool, that allows one to simulate hierarchically disordered systems, such as the ones which are used for the new generation thermoelectric materials. It has been proved successful in interpreting experiments [10, 11, 13, 15, 18, 24, 52 - 55], despite the fact that waves can only be approximated by particles only in the short wavelength limit, and despite the fact that long wavelength acoustic phonons carry a large portion of the heat. Thus, our aim is to quantify the margin of validity (or error) of the simple ray-tracing (which we know fails under certain conditions) compared to wave-based physics. Using the same simple geometries for both methods, we find that phonon transmission through the pores agrees very well for both wave-based and particle-based methods, with an error margin of ± 15%, across phonon wavelengths even for features with sizes down to 3-4 nm. Only for smaller feature sizes, the wave-based NEGF shows that long wavelength phonons can transmit easily, whereas we find that comparatively MC underestimates their transmission. In addition, particle-based simulation methods appear more sensitive to nanoscale structural variations compared to the wave-based NEGF method. Insights and knowledge gleaned from comparing wave and particle methods can be used to provide a better and more complete understanding of phonon transport in nanomaterials.

The paper is organized as follows: In Section II we describe the theoretical background and computational methods. In Section III we present our results, comparing wave-based Non-Equilibrium Green's Function (NEGF) and particle-based Monte Carlo ray-tracing simulations for simple geometries considering a single pore. In Section IV we discuss the importance of these results and examine geometries with more than one pore. Finally, in Section V we conclude.

## II.   Theoretical background and methods

The wave nature of phonons can be approximated by particles when they have wavelengths smaller than the characteristic length scales of the static disorder in the material under consideration (in our case pore size, pore distance, neck – see Fig. 1a). In materials with boundaries such as nanowires and thin films, when the boundary roughness amplitude in the structure is small, phonon reflection is primarily specular across wavevectors and correlation lengths [56]. For large roughness amplitudes, phonon reflections are diffusive, multiple scattering events at boundaries arise, and the thermal conductivity reduces significantly [11, 13, 55]. On the other side of the spectrum, long-wavelength phonons, much longer than the characteristic disorder in the nanostructure, cannot be effectively treated as particles, as at this point diffraction and interference effects become important. For example, Maurer *et* al. showed that such modes can be thought of as elastic-waves, which in the case of a large boundary roughness amplitude can experience a degree of localization, and Rayleigh waves can also appear in the material [46, 56].

Despite the limitation of Monte Carlo and ray-tracing when it comes to long wavelength phonons, the acoustic phonons that carry a significant portion of the heat are routinely treated as particles within Monte Carlo to describe thermal transport in nanostructures. Measured data has been successfully described in a large number of instances and materials with characteristic sizes smaller than the dominant phonon wavelengths [10, 11, 13, 15, 18, 24, 52 –54]. There is a growing interest in understanding thermal transport in hierarchically nanostructured materials, in which nanoscale features of various scales are embedded within a matrix material. In this case, Monte Carlo simulations are highly applicable, as they allow flexibility in the simulation domain geometry and size, and provide relatively good accuracy with reasonable computational cost. While large domains, larger than all important phonon wavelengths and mean-free-paths, can justify the treatment of the wave-phonons in terms of particles, still, in the vicinity of the smaller, nanoscale features, wave effects could persist. The influence of these features on the thermal conductivity under the wave versus particle phonon description is unknown.

The basis porous geometry we begin with is shown in Fig. 1a. The simple structure is defined by the pore diameter $D$ and the neck size $n$, which is the minimum distance between the pore boundary and the upper/lower boundaries. For computational efficiency we consider a basis two-dimensional (2D) simulation domain of length $L_x$ = 100 nm (x-direction) and width $L_y$ = 10 nm (y-direction). For all geometries we considered, we calculate the transmission of phonons from one side to the other for both methods: the wave-based NEGF method and the particle-based MC method.

<u>Non-Equilibrium Green's Function method:</u> The wave-based NEGF is a well-established, fully quantum mechanical method which can take into account the exact nanostructure geometry without any underlying assumptions. Thus, coherent effects are naturally captured. The method has been used primarily for electronic transport [57], but also for phonon transport in low-dimensional materials [58 - 64], yielding results in agreement with experimental measurements [59, 58].

It involves building the Dynamical matrix, which in our case is built atomistically using force constants [65]. A first nearest-neighbor force constant method is used to set up the dynamical matrix component between the $i^{th}$ and the $j^{th}$ silicon atoms, which are the first nearest-neighbors of each other. The force constant tensor is given by:

$$K_0^{(ij)} = \begin{bmatrix} f_r & 0 & 0 \\ 0 & f_{ti} & 0 \\ 0 & 0 & f_{to} \end{bmatrix} \quad (1)$$

where we use $f_r$ = 15.1319 N/m, $f_{ti}$ = 127.4988 N/m, and $f_{to}$ = 15.1319 N/m for the force constant fitting parameters we have chosen. Using these values the phononic bandstructure of bulk silicon is obtained to be in a relatively good agreement with the one we have calculated using the more complete, but more computationally expensive modified valence force field method in our previous works [59, 66]. The 3x3 dynamical matrix components are then calculated as:

$$D_{ij} = U_2^{-1} U_1^{-1} K_0^{(ij)} U_1 U_2 \qquad (2)$$

where $U_{1,2}$ are two unitary rotation matrices defined as:

$$U_1 = \begin{bmatrix} 1 & 0 & 0 \\ 0 & \cos\theta_{ij} & \sin\theta_{ij} \\ 0 & -\sin\theta_{ij} & \cos\theta_{ij} \end{bmatrix} \qquad (3)$$

$$U_2 = \begin{bmatrix} \cos\phi_{ij} & \sin\phi_{ij} & 0 \\ -\sin\phi_{ij} & \cos\phi_{ij} & 0 \\ 0 & 0 & 1 \end{bmatrix} \qquad (4)$$

where $q_{ij}$ and $\phi_{ij}$ represents, respectively, the polar and azimuthal angles of the bond between the $i^{th}$ and the $j^{th}$ silicon atoms. The dynamical matrix is then described as:

$$D = [D_{3\times3}^{(ij)}] = \frac{1}{M}\begin{Bmatrix} D_{ij} & i \neq j \\ -\sum_{l \neq i} D_{il} & i = j \end{Bmatrix} \qquad (5)$$

where $M$ is the silicon atomic mass. Using the device dynamical matrix, the Green's function is given by [57]:

$$G(E) = \left[(\hbar\omega)^2 I - D - \Sigma_1 - \Sigma_2\right]^{-1} \qquad (6)$$

here, the self-energy matrices for the contacts, $\Sigma_{1,2}$, are calculated using the Sancho-Rubio iterative scheme [67]. Afterwards, the ballistic transmission function can be computed using the relation:

$$T_{ph}(\omega) = Trace\left[\Gamma_1 G \Gamma_2 G^+\right] \qquad (7)$$

where, $\Gamma_{1,2} = i\left[\Sigma_{1,2} - \Sigma_{1,2}^+\right]$ are the broadening functions of the two contacts. The thermal conductance can then be obtained using the Landauer formula:

$$K_l = \frac{1}{2\pi\hbar} \int_0^\infty T_{\text{ph}}(\omega) \hbar\omega \left( \frac{\partial n(\omega)}{\partial T} \right) d(\hbar\omega) \tag{8}$$

where $n(w)$ is the Bose-Einstein distribution and $T$ is the temperature. Figure 1a shows typical channels we simulate, with one pore located in the middle. The channels we simulate with NEGF have length $L_x = 100$ nm, width $L_y = 10$ nm, but also a small thickness of 1 nm, as the Dynamical matrix is built on the atomistic lattice. The simulation domain contains 73,000 atoms.

Figure 1b depicts the phonon transmissions obtained with NEGF ($T_{\text{NEGF}}$) *versus* the energy ($\hbar\omega$) for the different geometries simulated. Transmission for the pristine case is given by the black line, followed by transmission for the different porous cases. On average, we observe that the transmission is reduced from the pristine channel as the pore size increases - in the figure we show the geometry with $D = 1$ nm (purple line), $D = 2$ nm (blue line), $D = 3$ nm (dark green line), $D = 4$ nm (light green line), $D = 5$ nm (orange line), and $D = 6$ nm (red line).

The typical phonon spectrum for a [100] Si nanowire of 10 nm width and 1 nm thickness is also calculated by applying periodic boundary conditions on its unit cell in the transport direction and shown in Fig. 1c, where the quantized phonon sub-bands are evident. The transmission of the pristine channel (black line in Fig. 1b) is essentially a count of the number of modes at each energy of Fig. 1c. For reference, in Fig. 1d we show the contribution of each phonon state to the total ballistic thermal conductance at room temperature calculated as:

$$K_l = \frac{k_B}{A} \sum_{i,q} v_{g,i}(q) \left[ \frac{\hbar\omega_i(q)}{k_B T} \right]^2 \frac{e^{\hbar\omega_i(q)/k_B T}}{\left( e^{\hbar\omega_i(q)/k_B T} - 1 \right)^2} \tag{9}$$

where $v_{g,i}(q)$ is the group velocity of a phonon with wavevector '$q$' in the '$i^{th}$' band and $A$ is the cross section area. Red and blue colors indicate the largest contribution and the smallest contribution, respectively (colormap). As seen, in addition to the low frequencies, the high frequency phonons have some contribution to the ballistic conduction as well.

In practice, however, Umklapp scattering is what limits the intrinsic conductivity, which also results in the high frequency phonons contributing only little to heat conduction.

The largest contribution is attributed to the low frequency acoustic modes. Recently, approaches based on the quantum mechanical NEGF method, using atomistic meshes and including anharmonic phonon-phonon scattering have been developed [68], however, these are bound to be computationally costly, and thus applicable to much smaller systems. Here, however, we seek something simpler: to isolate the coherent wave effects versus particle ray-tracing effects.

Monte Carlo ray-tracing method: For the particle-based simulations, we employ the ray-tracing Monte Carlo (MC) method to trace the phonon transmission in the nanostructures under consideration. Typically, Monte Carlo can accurately capture geometry details within micro–mesoscale domains and because of this it is widely employed to understand phonon transport in nanostructures [14 - 18, 20, 22 - 27, 29 - 43]. However, as we are only interested in the ballistic transmission here (where scattering happens only at the static pores), and in this case all phonons move in straight lines independent of their frequency, we follow a simpler version of what is usually employed in advanced MC simulations. In our MC method, phonons are initialized one at a time at the left contact with only a random initial angle, (as in the single-phonon incident-flux method [69, 70]). We do not use a specific dispersion relation, or the Bose Einstein distribution to initialize the population as all phonons across the dispersion are treated in the same way. Phonons are initialized on the left contact and allowed to travel through the simulation domain where they are either transmitted to the right or backscattered to the left. Once the phonon exits, the next phonon is then initialized. Thus, we only trace the phonon paths, and compute the transmission probability by taking a large number of phonon counts - ten million phonons are initialized on the left contact for each structure. The transmission from MC, unlike the one from NEGF, it is frequency independent, and essentially only depends on geometry. Hence for every structure considered, a single value of phonon transmission is determined for the MC cases. The domain discretization is 0.1 nm, and scattering on pores and boundaries is considered specular (the angle of incidence is the same as the angle of reflection).

Thus, for a pristine system, the MC transmission, $T_{P, MC}$ is 100%, i.e., all phonons that enter the system from the left contact leave from the right. When nanostructures are introduced in the domain, phonons backscatter, and the transmission, $T_{N, MC}$, is reduced from 100% depending on the pore diameter, $D$, and neck size $n$ (Fig. 1a). This transmission is normalized by the pristine value to get a "fractional transmission from pristine" value for the MC. This is given by:

$$F_{MC} = \frac{T_{N, MC}}{T_{P, MC}} \tag{10}$$

In the pristine case $F_{MC}$ is 1. Since scattering in MC is completely specular there is no wavevector (or frequency) dependence in the $F_{MC}$ (these are shown as solid lines in all figures from Fig. 2 onwards). In order to directly compare the MC transmissions to the NEGF transmissions, the latter are also converted into a normalized, fractional transmission in the same way:

$$F_{NEGF}(\omega) = \frac{T_{N, NEGF}(\omega)}{T_{P, NEGF}(\omega)} \tag{11}$$

However, since the NEGF transmissions have a frequency dependence, $F_{NEGF}$ also varies when plotted versus energy, $\hbar\omega$, because different energies behave differently in the presence of nanostructuring.

Note that although MC (ray-tracing) is completely particle-based, boundary scattering can be wavevector dependent and can be described using a boundary scattering treatment derived from diffraction theory [25, 55, 56, 71]. In all these cases, however, the scattering behaviour at a boundary depends on the boundary roughness strength, but here we consider the pore and the domain boundaries in both NEGF and MC to be completely smooth. Although scattering by the nanopore boundary is wavevector dependent within the NEGF formalism, we have no way to include this in ray-tracing without the need to assume the surface roughness amplitude. This would then lead to a situation of not being able to compare the two approaches on the same structure. Since we do not include the coherence-breaking [71 - 73] phonon-phonon scattering in NEGF (that would be computationally extremely demanding), we do not include it in MC either for one-to-one comparison. Thus, we investigate the purely coherent wave effects extracted from NEGF, to the ray-tracing MC in which case particles travel ballistically in the simulation domain, and the only source of scattering is the static disorder from the pores.

## III. Results

In order to evaluate the effect that each pore size (and corresponding neck) exert on the transmission, in Fig. 2 we consider the fractional transmissions, i.e., the ratios between each of the porous geometries to the pristine one. For this, the transmissions in all the porous structures (colored lines in Fig. 1b) are divided by the transmission of the pristine case (black line in Fig. 1b). For the case of NEGF transmissions, this is given by Eq. 11, and is plotted in Fig. 2 versus energy, $\hbar\omega$ (the color scheme is the same as in Fig. 1b). For every pore size the neck to diameter ratio ($n/D$) is also denoted on the right-side axis. The neck size characterizes the region available for phonon propagation, in addition to information about the scattering surface area which is implicit in the pore diameter. As we show later on, the $n/D$ ratio is a more accurate measure of the scattering strength compared to the diameter, or the neck alone. In Fig. 2, the average fractional transmission (i.e., the average of all the dots for each geometry), which we label $<T_{NEGF}>$, decreases from ~ 0.79 for the smallest pore size ($D$ = 1 nm, $n/D$ = 4.5, purple dots) to ~ 0.3 for the largest pore size ($D$ = 6 nm, $n/D$ = 0.33, red dots), a reduction of ~ 63%. Interestingly, if we consider only very the low frequency (long wavelength) acoustic phonons with energies under 3 meV, we see that the corresponding reduction is only ~ 10%. The high transmissions obtained with NEGF at low frequencies can be interpreted based on the wave-nature of phonons. Even for small neck sizes, low frequency, long wavelength phonons are more likely not to "see" the pores and propagate through/around them.

MC transmissions versus geometry: The exact same structures used for the NEGF calculations were also simulated with Monte Carlo. In Fig. 2, the solid straight lines indicate the MC fractional transmissions, given by Eq. 10, following the same coloring scheme for each corresponding structure. The average fractional transmission decreases from ~ 0.83 for the smallest pore size (purple line) to ~ 0.26 for the largest pore size (red line) a reduction of ~ 70% as in the case of NEGF. Comparing the flat MC lines with the NEGF dots, we can clearly see that MC provides a good approximation of the wave-based NEGF results for: i) all structures for energies above 10 meV (within an error margin to be discussed later in this section), and ii) for some of the smaller diameter structures with neck sizes above $n$ > 2 nm for all frequencies. This observation is quite interesting, as it quantifies the validity of particle-based methods for the case of long wavelength phonons, which is assumed to be problematic. It essentially shows that particle-based methods fail to describe the phonon transmission only when phonons are forced into extremely narrow constrictions of a few nanometres thick. In all other cases they seem to perform adequately.

Influence of neck size *n* and *n*/*D* ratio: In order to identify which structural feature size mostly affects the MC results, in Fig. 3 we compare the MC results in terms of the neck size *n*, diameter *D*, and their *n*/*D* ratios. Here, each line plots the fractional transmissions for each porous structure of a given pore diameter using the same coloring as earlier. For each pore size the neck is increased by increasing the width of the simulation domain $L_y$. Typical geometries simulated are given in Fig. 3 (I – IV). Uniform *n*/*D* values are indicated by the dashed black lines. For a fixed diameter, the transmission increases as the neck size increases, as expected. In the case of *D* = 6 nm (red line) the smallest neck size, *n* = 2 nm (geometry IV in panel above), yields the smallest transmission, and increasing the neck size for the same pore size (moving up the red line) increases fractional transmission to ~ 0.83 (geometry III in panel above). The last point of the red line corresponds to *n*/*D* = 4.5. Interestingly, all other structures of different diameters, but of the same *n/D*, have the same fractional transmission (horizontal dashed lines for constant *n*/*D* ratios across Fig. 3). For the same neck size, on the other hand, as the diameter increases (moving vertically downwards from line to line), the fractional transmission decreases. Thus, the *n*/*D* ratio is better correlated to the MC fractional transmissions in these structures, than either *D* or *n* alone, indicating a better metric for identifying the influence of porosity and constrictions together. We have verified the strong correlation between *n*/*D* and thermal transport in molecular dynamics (MD) simulations as well. We will discuss the effect of geometry based on this metric from here on.

Comparing NEGF and MC variations: We now quantify the variation between the fractional transmissions given by the two methods. In Fig. 4 we plot their variation as a percentage of the MC value ($F_{NEGF}/F_{MC}$ -1). For the MC this quantity is flat at zero for all structures, i.e., $F_{MC}/F_{MC}$ -1, and allows to extract a global variation measure between NEGF and MC across geometries and frequencies. The data in Fig. 4 uses the same coloring scheme as in Fig. 2. The *n*/*D* values indicated in each case range from *n*/*D* = 4.5 (purple dots) to *n*/*D* = 0.33 (red dots). Overall, the NEGF fractional transmissions oscillate around the MC fractional transmissions (at 0). The variation increases with decreasing *n*/*D* ratio (compare red *vs* purple dots for the smaller/larger *n/D*, respectively). Again, the variation is highest for low frequency (long wavelength) acoustic phonons with energies under 3 meV, where variation in excess of ~100% can be seen for *n/D* < 0.5, i.e., the MC underestimates significantly the NEGF at these energies. However, for phonons with energies greater than 8 meV, the MC method is able to approximate the NEGF phonon transmission within a 15% error margin especially for the smaller diameters/large neck sizes.

The 15% margin within which the MC is able to approximate the NEGF is a consequence of the frequency dependence of the NEGF transmissions. This results in variations around a mean value that is in any case adequately captured by the MC ray-tracing transmissions. In fact, this margin can also be captured in simple geometrical variations. We have simulated using MC structures in which we have altered the size of the pores, as seen in Fig. 5a and Fig. 5b such that *n/D* is varied by ±15% for the smaller and the larger diameter structures. Smaller/larger pores will allow larger/smaller transmissions. The dashed lines in Fig. 5c indicate the change in the MC transmission as the *n/D* changes. Effectively this encompasses a larger proportion of the frequency dependent NEGF transmissions, but still not for the low frequency, long wavelength phonons, especially in the smaller *n/D* structure.

## IV. Discussion

In the results above we see variations between the NEGF and the MC of 25% - 100% for frequencies below 8 meV, at low feature sizes, i.e., below *n/D* = 0.5, *n* = 2 nm. This quantifies the region of validity of particle-based approaches for this specific nanostructure (at least). However, looking at the global average of the NEGF transmissions (dots) we find that the error is less than 15% for most of the spectrum. This is shown in Fig. 6, which plots the global average transmission of the NEGF results for all frequencies, $<T_{\text{NEGF}}>$, at a given *n/D* ratio (dashed-red line), and compares them to the values of MC for those same structures (blue line). The error margin for all *n/D* is less than 15%. MC slightly overestimates the transmission for structures with large necks and small pore sizes (when *n/D* >2) where *D* is the limiting variable. It slightly underestimates the transmission for structures with small neck sizes and large pore sizes (*n/D* < 2), where *n* is the limiting variable. Thus, particle-based MC captures average wave-based thermal transmission effects largely within 15% down to constriction feature sizes of 2 nm. This knowledge could be very useful in quantifying the accuracy of simulations of thermal transport in nanostructured materials under different simulation methods.

In order to further quantify the upper limit of the error that could be expected in MC, in Fig. 6 we also look at $<T_{\text{NEGF}}>$ in a part of the spectrum up to 5 meV (purple dashed line) and up to 15 meV (green dashed line). The deviation from the MC blue line, and thus the potential error, is at a maximum for the small $n/D$ cases. At $n/D = 0.33$, the error is 120% when we consider $<T_{\text{NEGF}}>$ up to 5 meV. Conversely, the error is at its lowest (below 3%) for high $n/D$ structures (see $n/D = 4.5$). In this case, thermal transport properties predicted by MC would be very applicable for nanostructures with $n/D$ of 2 (or neck size 2 nm) and above. On the other hand, we find that the global average error (variation from MC value) for $<T_{\text{NEGF}}>$ up to 5 meV (purple dashed line) is 46% while, if frequencies up to 15 meV (green dashed line) are considered this reduces to 12%.

However, we must note that a major component of the heat in Si is carried by long wavelength acoustic phonons with frequencies below 20 meV [15, 52, 74 - 76]. This can result in the error in MC being larger in narrow constrictions, as it cannot capture the transmission of these long wavelength phonons accurately. On the other hand, in the presence of nanostructuring, where the phonon mean-free-paths are limited by the geometrical feature sizes, the relative contribution of phonons from the rest of the spectrum would increase [78 - 80]. This can be intuitively seen from Matthiessen's rule, in which the overall mean-free-path of a phonon is dominated by the stronger of the scattering processes. Thus, a long mean-free-path will suffer relatively more in the presence of nanostructuring compared to shorter MFP phonons, which reduces the gap between their relative contributions to heat current. In fact, it is pointed out that optical phonons and the rest of the spectrum can also carry a significant portion of the heat at the nanoscale because longer (acoustic) mean-free-path phonons are scattered more strongly leading to a proportionally larger contribution to transport from (optical) mean-free-paths [81].

Multi-pore systems: We have repeatedly seen in our results that MC underestimates transmissions for long wavelength phonons, compared to NEGF. This is due to the particle-nature of the method, in which by definition phonons always scatter when they hit pores. From a structural examination point of view, this also means the MC method is more sensitive to structural changes in the device. To study this more closely we look at structures with more than one pore. In Fig. 7 we simulate a system with two pores and measure the effect of pore separation on both the NEGF and the MC fractional transmissions. The pore diameter is kept constant at $D = 5$ nm and the pores are positioned equidistantly from the central point, with changing their horizontal pore separation length, $l$, along the $x$-axis. Taking the average $<T_{NEGF}>$ for all energies at each given $l$ we observe that the fractional transmission is reduced from an average of 0.27 in the $l = 0$ nm case (black dots) in Fig. 7, where the pore boundaries are just touching, to 0.25, where the pores are separated by $l = 25$ nm (red dots). This is a difference of ~8%. In the case of MC, we see a larger change in fractional transmissions for the same structures from 0.27 in the $l = 0$ nm case (black line) to 0.22 when the pores are separated by $l = 25$ nm (red line). This is a difference of 20%, as seen by comparing the two straight lines, more than double of what the NEGF experiences.

The case of a vertical separation along the $y$-axis, is explored in Fig. 8, with pore sizes fixed at $D = 2$ nm. The pores are placed at the mid-point of the structure along the $x$-axis and equidistantly from the central point in the $y$-axis, with changing their pore separation distance, $d$. Taking the average $<T_{NEGF}>$ for all energies for each $d$, we observe that the fractional transmission reduces from an average of 0.49 in the $d = 0$ nm case (black dots) to 0.47 in the $d = 3$ nm case (red dots). This is a difference of only 4%. In the case of MC, we see a larger change in fractional transmissions for the same structures from 0.5 in the $d = 0$ nm case (black line) to 0.45 in the $d = 3$ nm case (red line). This is a difference of 10%, again more than double compared to that observed in the NEGF case. Thus, we can conclude that the particle-based MC is more sensitive to geometrical details in the material structure than the wave-based NEGF at these feature sizes between 1-10 nm. Again, this is attributed to long wavelength phonons transmitting through or around the pores, being less affected by their structural details in NEGF. It is, however, possible to approximate the NEGF results by changing the effective size of the features simulated in MC, as we see in Fig. 5. For low feature sizes an effective decrease in pore size would increase transmission in MC to reduce the deviation from NEGF.

## V. Conclusions

In this work, we investigated the effect of the wave versus particle nature of phonons in their flow through nanostructured porous Si with small pores and constrictions. We used a full wave approach based on the atomistic Non-Equilibrium Green's Function (NEGF) method, and a particle-based approach using ray-tracing Monte Carlo (MC). Using the same simple geometries for both methods, we showed that phonon transmission through the pores agrees well for both methods, with an error margin of ± 15% across phonon wavelengths even for constriction sizes as small as 2 nm and pore diameters as small as 1 nm as well. We also show that the neck to diameter ratio, *n/D* is a better measure of the effect of geometry, rather than the neck, or the diameter alone. We find that MC significantly underestimates the transmission of long wavelength phonons only in structures with *n/D* < 2 compared to NEGF. Long wavelength phonons are shown to propagate more easily through small constrictions, compared to what a particle treatment would suggest. We also found that the particle-based MC is more sensitive to structural details compared to the wave-based NEGF method. Overall, this work suggests that in spite of the different assumptions made by each model, it is possible to use the MC even at the nanoscale and obtain results in agreement (within 15%) with NEGF, even down to very small features. Insights and features extracted from our comparison of wave versus particle methods can be useful in providing a better and more complete understanding of phonon transport in nanomaterials.

## VI. Acknowledgements

This work has received funding from the European Research Council (ERC) under the European Union's Horizon 2020 Research and Innovation Programme (Grant Agreement No. 678763).

Figure 1:

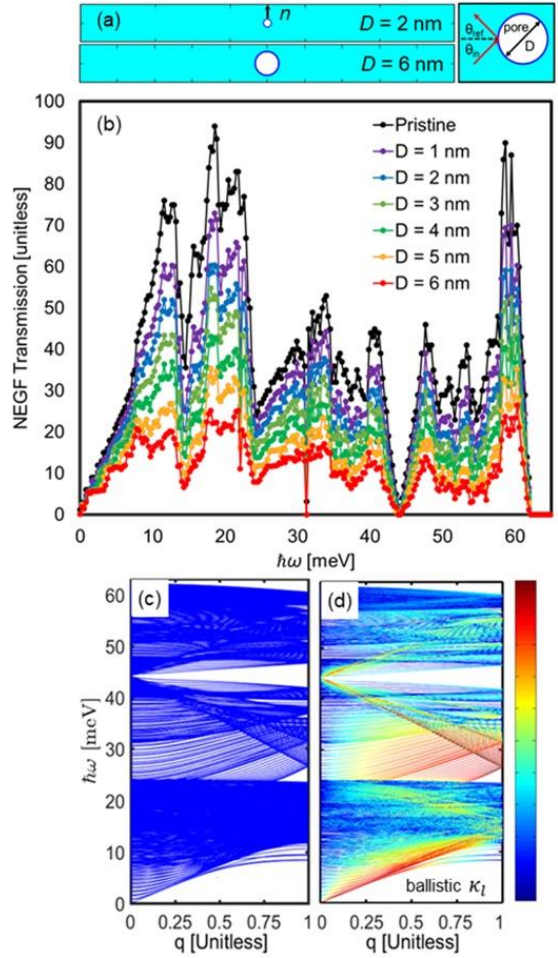

Figure 1 caption:

(a) Schematics of typical geometries studied with length of 100 nm, width 10 nm and thickness 1 nm for two pore diameters of $D = 2$ nm and $D = 6$ nm. The neck, $n$, is measured, as indicated, as the widest distance between the edge of the pore and the nearest geometry boundary. (b) The NEGF transmission [unitless] *vs* the energy, $\hbar\omega$. Transmission for the pristine case ($T_{P, NEGF}$) is given by the black line. Transmission for nanostructured porous cases ($T_{N, NEGF}$) are given for $D = 1$ nm (purple line), $D = 2$ nm (blue line), $D = 3$ nm (dark green line), $D = 4$ nm (light green line), $D = 5$ nm (orange line), $D = 6$ nm (red dots) respectively. (c) The typical phonon spectrum for a [100] Si nanowire of 10 nm width and 1 nm thickness calculated by applying periodic boundary conditions on the unit cell in the transport direction. The transmission of the pristine channel (black line in Fig. 1b) is essentially a count of the number of modes at each energy of Fig. 1c. (d) The contribution of each phonon state to the total ballistic thermal conductance at room temperature calculated using Eq. 8 (without Umpklapp scattering). Red and blue colors indicate the largest contribution and the smallest contribution, respectively (colormap).

Figure 2:

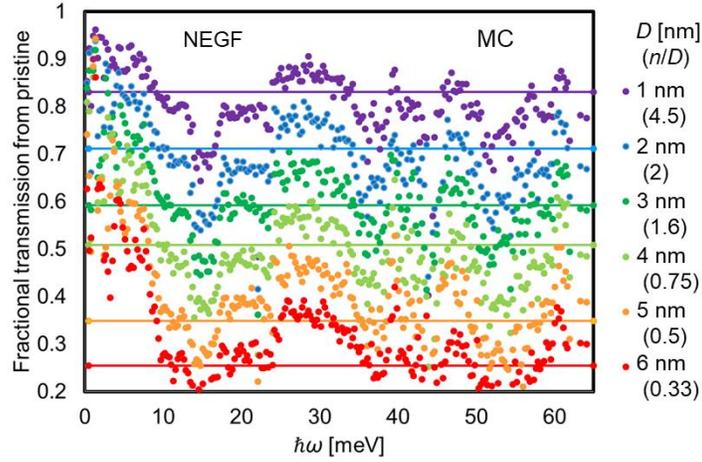

Figure 2 caption:

NEGF fractional transmission $F_{NEGF}$ versus energy $\hbar\omega$ given by Eq. 11, for $D = 1$ nm (purple dots), $D = 2$ nm (blue dots), $D = 3$ nm (dark green dots), $D = 4$ nm (light green dots), $D = 5$ nm (orange dots), $D = 6$ nm (red dots), respectively. Monte Carlo fractional transmission $F_{MC}$ given by Eq. 10, for the same structures. The pore diameters ($D$) and the corresponding neck to diameter ($n/D$) ratios are provided on the right axis.

Figure 3:

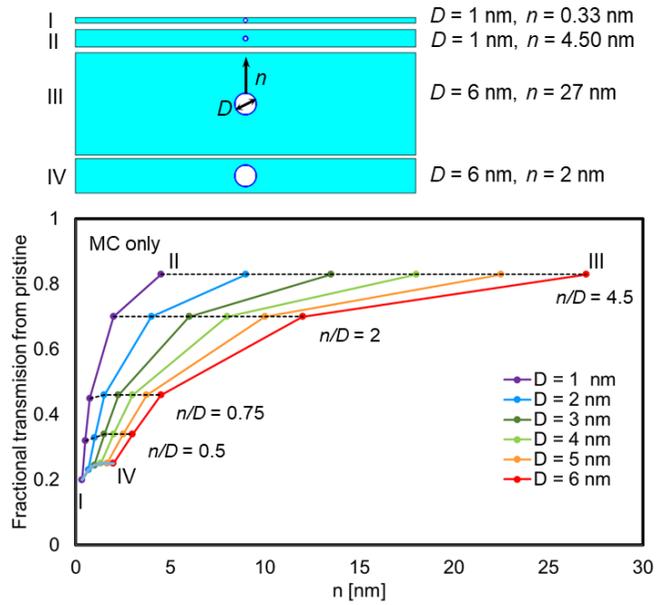

Figure 3 caption:

Monte Carlo fractional transmission $F_{MC}$ given by Eq. 10, for $D = 1$ nm (purple line), $D = 2$ nm (blue line), $D = 3$ nm (dark green line), $D = 4$ nm (light green line), $D = 5$ nm (orange line), $D = 6$ nm (red line), respectively, versus neck size, $n$. Typical geometries simulated are depicted in the panel above (I – IV). The common $n/D$ values across structures are indicated by dashed lines.

Figure 4:

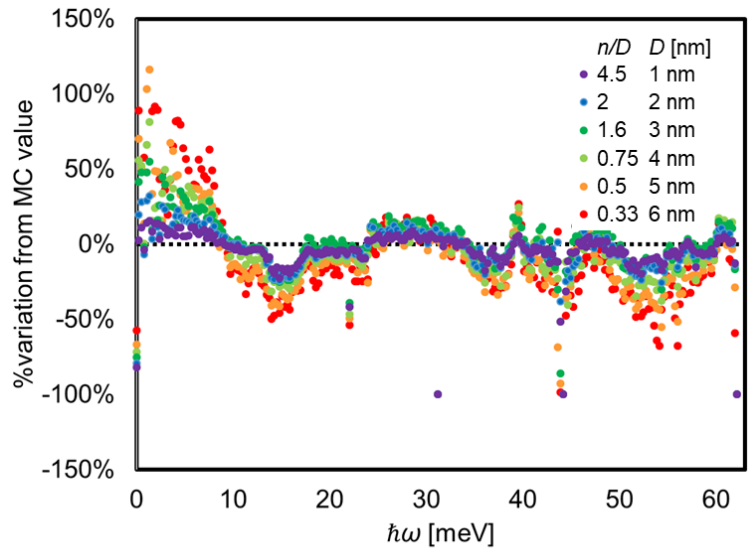

## Figure 4 caption

Percentage variation of the $F_{NEGF}$ from the $F_{MC}$ values for $D = 1$ nm (purple dots), $D = 2$ nm (blue dots), $D = 3$ nm (dark green dots), $D = 4$ nm (light green dots), $D = 5$ nm (orange dots), $D = 6$ nm (red dots), respectively, versus energy $\hbar\omega$.

Figure 5:

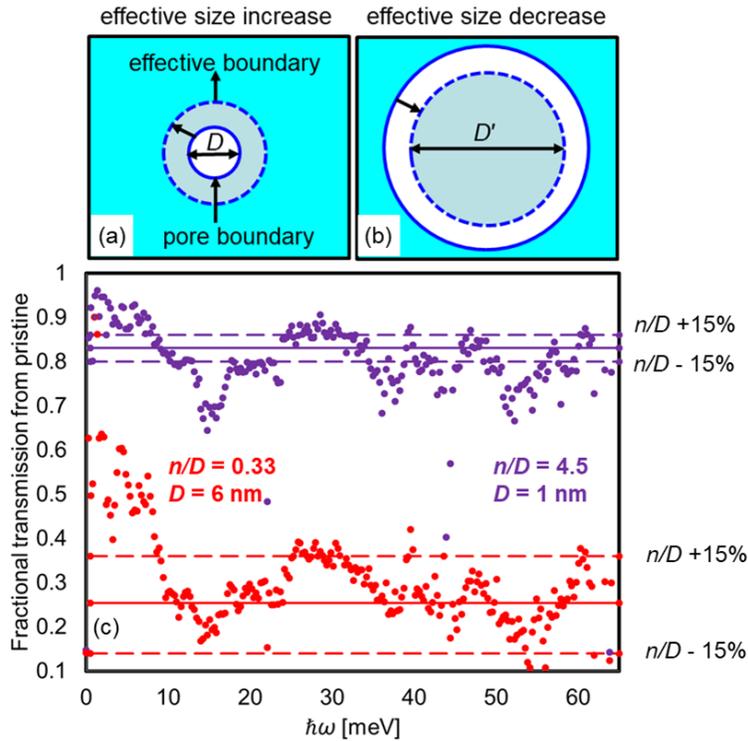

Figure 5 caption:

(a) Schematic of effective increase in *D* and (b) effective decrease in *D*. (c) NEGF fractional transmission for structures with *n*/*D* = 4.5, *D* = 1 nm (purple dots), and *n*/*D* = 0.33, *D* = 6 nm (red dots), respectively, versus energy *ℏω*. MC fractional transmission for the same structures are plotted by the solid lines. The dashed lines represent Monte Carlo ray-tracing for structures with a ± 15% variation on their *n*/*D* ratio.

Figure 6:

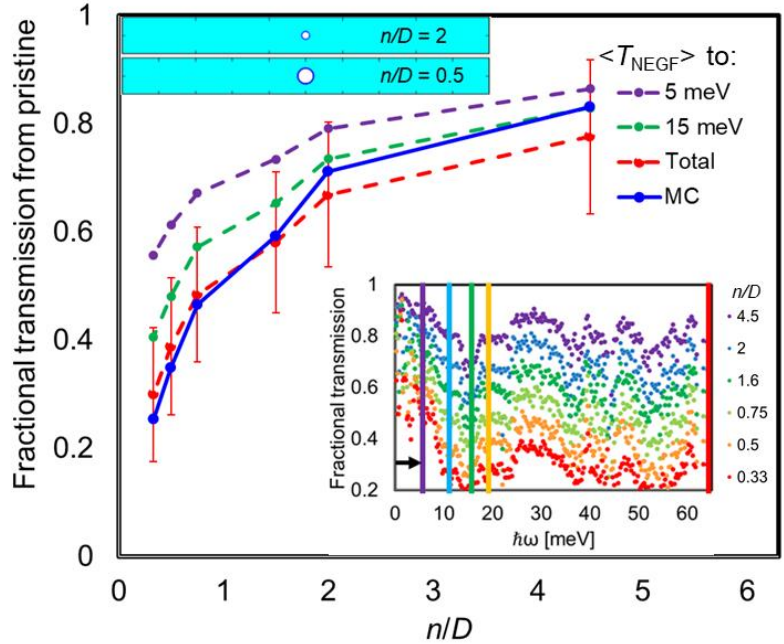

Figure 6 caption:

NEGF results averaged $<T_{NEGF}>$ over energy in a part of the spectrum up to 5 meV (purple dashed line), up to 15 meV (green dashed line) and for the total spectrum (red dashed line) *vs n/D*. These limits are depicted in the bottom inset with lines of corresponding colors. Error bars give the standard deviation of $T_{NEGF}$ data. MC fractional transmission *vs n/D* (blue solid line) is given for comparison. Schematics of some geometries simulated are depicted at the top left corner.

Figure 7:

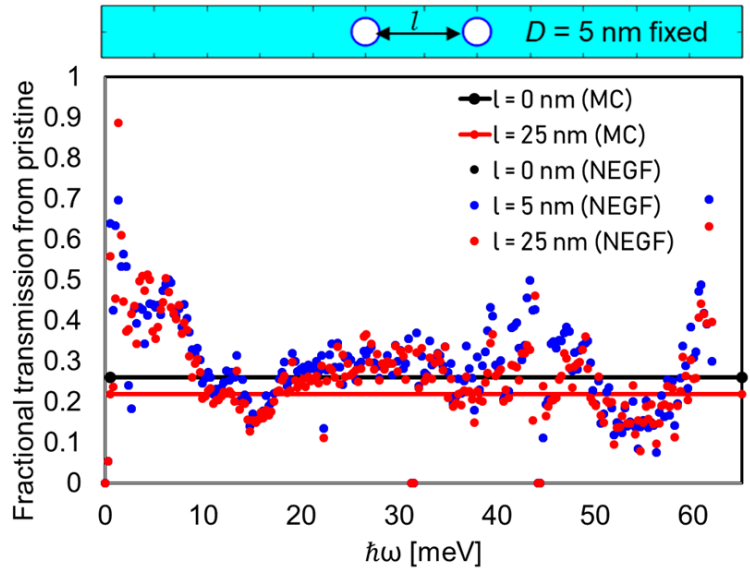

Figure 7 caption:

NEGF fractional transmission for two-pore structure with pore separation $l = 0$ nm (black dots), $l = 5$ nm (blue dots), $l = 25$ nm (red dots) versus energy. MC fractional transmission for the same structures is shown by the solid lines. Schematic of a typical structure simulated with $D = 5$ nm is given in the panel above.

Figure 8:

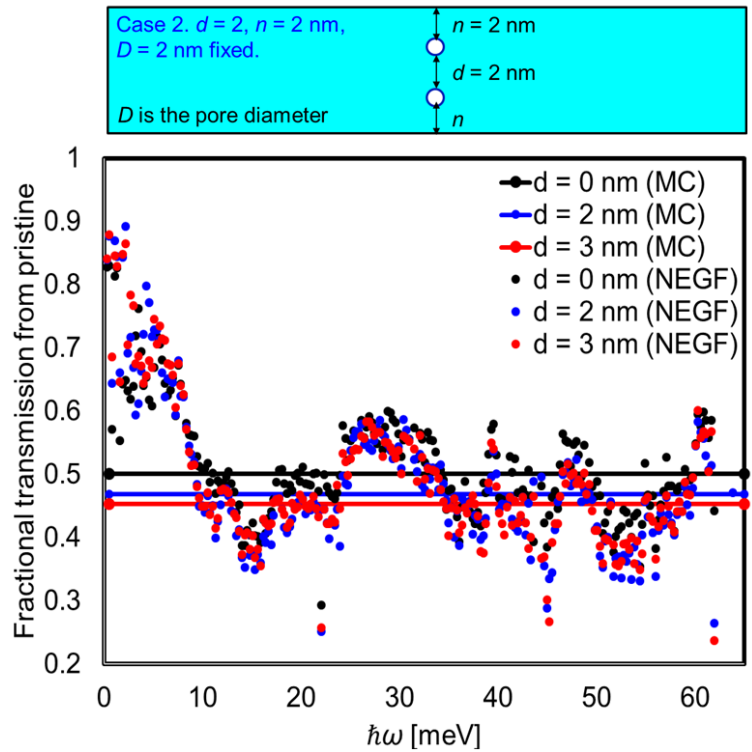

Figure 8 caption:

NEGF fractional transmission for two-pore structures with vertical separation $d = 0$ nm (black dots), $d = 2$ nm (blue dots), $d = 3$ nm (red dots) versus energy. MC fractional transmissions for the same strucutres are shown by the solid lines. A schematic of a typical structure simulated with $D = 2$ nm is given in the panel above.